\documentclass[twocolumn,showpacs,preprintnumbers,amsmath,amssymb,floatfix,superscriptaddress,prl]{revtex4}
\usepackage[dvips]{graphicx}% Include figure files
\usepackage{dcolumn}% Align table columns on decimal point
\usepackage{bm}% bold math
\usepackage[usenames]{color}
\newcommand{\md}[1]{}  % do not print  deletions
\newcommand{\nyy}{n_{YY}}  % Yang Yang  density
\begin{document}

\preprint{}

\title{Yang-Yang thermodynamics on an atom chip}

\author{A.~H. van Amerongen}
\author{J.~J.~P. van Es}
\author{P. Wicke}
\affiliation{Van der Waals-Zeeman Institute, University of Amsterdam,
Valckenierstraat 65-67, 1018 XE Amsterdam, The Netherlands}
\author{K.~V. Kheruntsyan}
\affiliation{ARC Centre of Excellence for Quantum-Atom Optics, School of Physical
Sciences, University of Queensland, Brisbane, Queensland 4072, Australia}
\author{N.~J. van Druten}
%\email{druten@science.uva.nl}
%\homepage{http://www.science.uva.nl/research/aplp/}
\affiliation{Van der Waals-Zeeman Institute, University of Amsterdam,
Valckenierstraat 65-67, 1018 XE Amsterdam, The Netherlands}
\date{\today}% It is always \today, today,
             %  but any date may be explicitly specified

\begin{abstract}
We investigate the behavior of a weakly interacting nearly
one-dimensional (1D) trapped Bose gas at finite temperature. We perform
\emph{in situ} measurements of spatial density profiles and show that
they are very well described by a model based on exact solutions
obtained using the Yang-Yang thermodynamic formalism, in a regime where
other, approximate theoretical approaches fail. We use
 Bose-gas focusing [Shvarchuck {\it et~al.}, 
Phys. Rev. Lett. {\bf 89}, 270404 (2002)]
to probe the axial momentum
distribution of the gas, and find good agreement with the \emph{in
situ} results.
\end{abstract}

\pacs{03.75.Hh, 05.30.Jp, 05.70.Ce}

\maketitle

Reducing the dimensionality in a quantum system can have dramatic
consequences. For example, the 1D Bose gas with repulsive
delta-function interaction exhibits a surprisingly rich variety of
physical regimes that is not present in 2D or 3D
\cite{PetGanShl04,Castin04}. This 1D Bose gas model is of particular
interest because exact solutions for the many-body eigenstates can be
obtained using a Bethe ansatz \cite{LieLin63}.
 Furthermore, the
finite-temperature equilibrium can be studied using the Yang-Yang
thermodynamic formalism \cite{YanYan69,KorBogIze93,Tak99}, a method
also known as the thermodynamic Bethe ansatz. This formalism is the
unifying framework for the thermodynamics of a wide range of exactly
solvable models. It yields solutions to a number of
important interacting many-body quantum systems and as such
provides critical benchmarks to condensed-matter physics and
field theory \cite{Tak99}. The specific case of the 1D Bose gas as
originally solved by Yang and Yang \cite{YanYan69} is of particular
interest because it is the simplest example of the formalism.
 The experimental
achievement of ultracold atomic Bose gases in the 1D regime
\cite{GorVogLea01andothers} has attracted renewed attention to the 1D
Bose gas problem \cite{1D-theory-others} and is now providing
previously unattainable opportunities to test the Yang-Yang
thermodynamics.

In this paper, we present the first direct comparison between
experiments and theory based on the Yang-Yang exact solutions. The
comparison is done in the weakly interacting regime and covers a wide
parameter range where conventional models fail to quantitatively
describe \textit{in situ} measured spatial density profiles.
Furthermore, we use Bose-gas focusing \cite{ShvBugPet02}
to probe the equilibrium momentum distribution of the 1D gas, which is
difficult to obtain through other means.

For a uniform 1D Bose gas, the key parameter is the dimensionless
interaction strength $\gamma=mg/\hbar^2 n$, where $m$ is the mass of
the particles, $n$ is the 1D density, and $g$ is the 1D coupling
constant. At low densities or large coupling strength such that
$\gamma\gg 1$, the gas is in the strongly interacting or
Tonks-Girardeau regime~\cite{Gir60KinWenWei04ParWidMur04}.~The
opposite limit $\gamma\ll 1$ corresponds to the weakly interacting
gas. Here, for temperatures below the degeneracy temperature
$T_d=\hbar^2 n^2/2 m k_B$, one distinguishes two regimes
\cite{KheGanDru03}. \textit{(i)} For sufficiently low temperatures,
$T\ll \sqrt{\gamma}T_{d}$, the equilibrium state is a
quasi-condensate with suppressed density fluctuations. The system
can be treated by the mean-field approach and by the Bogoliubov
theory of excitations. The 1D character manifests itself through
long-wavelength phase fluctuations resulting in a finite phase
coherence length $l_\phi$ which greatly
exceeds the mean-field correlation length $l_c$. \textit{(ii)} The
temperature interval $\sqrt{\gamma}T_{d}\ll T\ll T_{d}$ corresponds
to the quantum decoherent regime \cite{KheGanDru03}, where both the
density and the phase fluctuate. Here, the condition $l_{c}\ll
l_{\phi}$ required for the existence of a quasi-condensate is no
longer satisfied and the system can be treated as a degenerate ideal
gas combined with perturbation theory in $g$. At temperatures near
the crossover to the quasi-condensate, $T\sim \sqrt{\gamma}T_{d}$,
neither of the approximate 
approaches work and one has to rely on the numerical solution to the
exact Yang-Yang equations, as we show in the present work.

Experiments on 1D gases are usually carried out in
harmonic traps with strong transverse confinement and weak
confinement along the axis, $\omega_\perp\gg
\omega_\|$. A trapped gas is in the 1D regime if both temperature
and chemical potential are small with respect to the radial
excitation energy, $k_{B}T,\mu\ll\hbar\omega_\perp$. It was
recognized early on that the physics of the degenerate part of the
trapped cloud is already effectively 1D if the weaker condition
$\mu<\hbar \omega_\perp$ is satisfied \cite{GorVogLea01andothers}.
For a gas in the 1D regime, the effective 1D coupling
can be expressed through the 3D scattering length $a$ as $g\simeq 2
\hbar\omega_\perp a$ if $a\ll (\hbar/m \omega_\perp)^{1/2}$
\cite{Ols98}.

Various physical regimes of a harmonically trapped 1D gas
have been discussed in
Refs.~\cite{PetShlWal00,PetGanShl04,KheGanDru05,BouKheShl07}.
 The above
classification of the regimes for the uniform gas can be applied
locally to the trapped gas if the conditions for the local density
approximation (LDA) are met
\cite{PetGanShl04,KheGanDru05,BouKheShl07}. The trapped cloud is
then characterized by a global temperature and a spatially varying
density, $n\rightarrow n(x)$. Thus, a trapped 1D gas which is in
the quasi-condensate regime in its centre [$T\ll
\sqrt{\gamma(0)}T_d(0)$] will cross over towards the wings --
through the quantum-decoherent regime -- to the nondegenerate regime
[$T_d(x)\ll T$] as the local density $n(x)$ decreases.

We experimentally investigate the behavior of a weakly interacting
trapped Bose gas [$\gamma(0)\simeq 10^{-2}$] in the regime
where $\mu<\hbar\omega_{\perp}$ and $k_BT \simeq
\hbar\omega_{\perp}$. Similar measurements to our {\em in situ}
data were previously performed at higher temperatures and higher
linear densities \cite{EstTreSch06,TreEstWes06}; the observed density profiles were
found to be in disagreement with approximate theories. Our approach
here is different in that we present a full description of the
density profiles through a consistent model based on the solutions
to the exact Yang-Yang equations \cite{YanYan69,KheGanDru03} and
that we compare real-space and momentum-space distributions for the
same experimental parameters.

For (nearly) 1D clouds, it is difficult to probe the axial momentum
distribution. The conventional time-of-flight method does not work,
mainly because the cloud hardly expands axially beyond its initial
length. The analysis is further complicated by strong density
fluctuations that develop in time-of-flight from the initial phase
fluctuations \cite{comment1}. The axial momentum distribution of 
phase-fluctuating 3D condensates has been measured using Bragg 
spectroscopy \cite{Ric03}, but this technique is extremely sensitive to 
vibrations and requires averaging over many realizations.

We gain experimental access to the axial momentum distribution using
Bose gas focusing. This is a technique introduced for
non-equilibrium 3D condensates \cite{ShvBugPet02}; here, we extend
it to study the equilibrium properties of (nearly) 1D clouds. A
detailed analysis of focusing in this regime will be presented
elsewhere. In brief, we apply a short, strong axial harmonic
potential yielding a kick to the atoms proportional to their
distance from the trap center
(analogous to the action of a lens in optics),
followed by free propagation. As a result
the atoms come to a focus, at which time the axial density distribution
reflects the axial momentum distribution before focusing. Initial phase
fluctuations result in a finite width of the cloud \cite{ShvBugPet02}. Since
the focusing brings all atoms together axially, the signal level is
high, even for a single realization. As we will show, averaging over a
few shots is sufficient to obtain high signal-to-noise ratio.

The core of our setup is a magnetic microtrap
consisting of three layers of current carrying wires. The surface
layer is formed by a silicon substrate coated with a
1.8-$\mu$m-thick vapor-deposited patterned gold layer. This atom
chip faces in the direction of gravity ($+z$). On the chip we use a
Z-shaped wire, with a 3~mm long and $125~\mu$m wide central section
along $x$. The remaining two layers are behind the silicon substrate
and each contain three parallel copper wires (diameter
300~$\mu$m), in the $x$ and $y$-direction, centered at $z=-0.5$~mm
and $-0.8$~mm from the chip surface, respectively. The wires in
the $y$ direction are used to vary the confinement along $x$.
\begin{figure}[tbp]
\includegraphics[width=85mm]{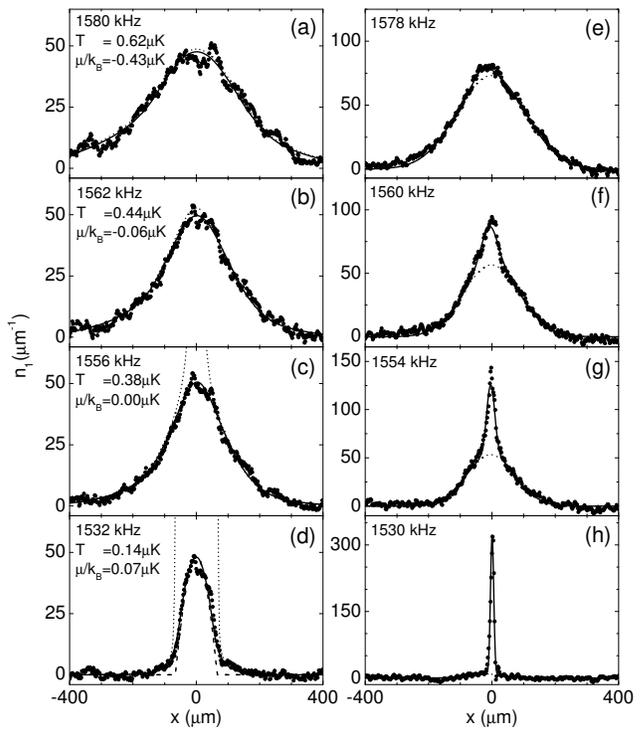}%130mm
\caption{Linear atomic density from absorption images obtained
\textit{in situ} (a)-(d) and \textit{in focus} (e)-(h) by
lowering (from top to bottom as indicated) the final RF evaporation
frequency. \textit{In situ}: solid lines are fits using Yang-Yang
thermodynamic equations (see text). The values of $\protect \mu$ and
$T$ resulting from the fits are shown in the figure. Dotted line:
ideal Bose gas profile showing divergence for $\protect\mu(x)=0$.
Dashed line in (d): quasi-condensate profile with the same peak
density as the experimental data. \textit{In focus}: solid lines are
the sum of two independent Gaussian fits -- one to the wings (dotted
lines) and one to the central part of the density profile.}
\label{fig:SituFocus}
\end{figure}

We trap $2\times 10^7$  $^{87}$Rb atoms in the $F=2, m_F=2$ state in
a tight magnetic trap near the chip surface, and perform forced
evaporative cooling by applying a radio frequency (RF) field. The
frequency $\omega_{RF}$ is ramped down from $27~$MHz to $1.7~$MHz
relatively quickly (in 180~ms) to purposely reduce the atom number,
in order to penetrate deeper into the 1D regime. Before reaching
degeneracy we relax the axial confinement to a final trap with
$\omega_{\perp}/2\pi=3280$~Hz, $~\omega_{\|}/2\pi=8.5$~Hz, and a
bottom corresponding to $\omega_{RF}/2\pi=1.518(2)$~MHz. The current
in the Z-wire is set at 2.25~A, and the distance of the cloud to the
chip surface is 90~$\mu$m. In this trap we perform a slower ramp
(450~ms) to the final RF frequency. An additional 300~ms of plain
evaporation allows the damping of residual quadrupole collective
oscillations in the cloud to the point where these oscillations are
no longer visible. We probe the gas using standard absorption
imaging, with a measured optical resolution limit of $4~\mu$m. The CCD camera
square pixel size is $ 2.15~\mu$m in the object plane.
Our axial trapping potential was characterized using both the
measurement of \textit{in situ} density profiles at high $T$
and the dipole mode oscillation frequency
in the trap center. The curvature in the center corresponds to a
frequency of $8.5$~Hz and gradually decreases to $\sim6$~Hz in the
wings of the cloud \cite{potspec}.

In Fig.~\ref{fig:SituFocus}(a)-(d) we show the linear density of atomic
clouds in the magnetic trap for different final RF frequencies. These
data were obtained by \textit{in situ} absorption imaging and
integrating the atom number along $z$. The absolute atom number was
calibrated using time-of-flight data. Each curve is an average of $\sim
18$ images taken under identical circumstances. Since all of our data
was taken for $\mu<\hbar\omega_\perp$ \cite{comment2}, we expect that
the interactions will significantly affect only the distribution in the
radial \emph{ground} state, while the population in the radially
\emph{excited} states can to a good approximation be described by the
ideal-gas distribution. This leads to the following model that was used
to analyze the \textit{in situ} data.

\begin{figure}[tbp]
\includegraphics[width=77mm]{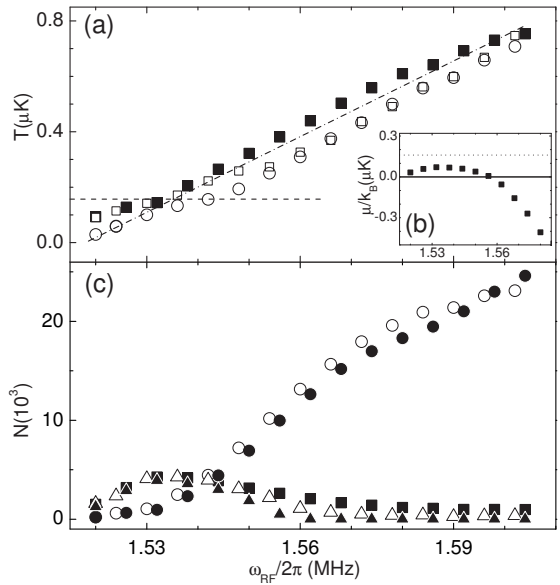}%110mm
\caption{Characterization of the atomic clouds
as a function of the final RF frequency, as determined from fits of the Yang-Yang
model to the {\em in situ} data and Gaussian fits to the {\em in
focus} data. (a) Temperature from the {\em in situ}
data ($\blacksquare$) and from the radial ($\square$) and axial
($\circ$) size of the broad Gaussian fit to the {\em in focus}
data. The dash-dotted line is to guide the eye and indicates a ratio
of 11 of the trap depth and the
cloud temperature; dashed line corresponds to $\hbar
\protect\omega_{\perp}/k_B$. (b) Chemical potential from the
Yang-Yang fit; dashed line indicates $\hbar \protect\omega
_{\perp}/k_B$. (c) Atom number from the {\em in focus}
data:  wide distribution ($\circ$) and central peak  ($\triangle$); from
the Yang-Yang model fit to the {\em in situ} data:
atoms in the radial ground state ($\blacksquare$), in radially
excited states ($\bullet$), and atoms in the radial ground state
experiencing $\mu(x)>0$ ($\blacktriangle $). }
\label{fig:temperature}
\end{figure}

We start from the solution to the Yang-Yang integral equations for a
finite-temperature uniform 1D Bose gas at thermal equilibrium
\cite{YanYan69}. This yields numerical results for both the equation
of state $\nyy(\mu,T)$ and the local pair correlation 
\cite{KheGanDru03}. The LDA is then used to account for
the axial potential via a varying chemical potential
$\mu(x)=\mu-V(x)$. This approach is expected to be valid as long as
the axial potential is smooth on the scale of the relevant
correlation lengths \cite{KheGanDru05,BouKheShl07}. Since our
temperature is on the order of the radial level splitting,
$\hbar\omega_\perp/k_B=158$~nK, the fraction of the gas which
occupies radially excited states can not be neglected. We account
for this fraction by summing over radially excited states (radial
quantum number $j\geq 1$, degeneracy $j+1$) and treating each radial
state as an independent ideal 1D Bose gas in thermal equilibrium
with the gas in the radial ground state,
$\mu_j(x)=\mu(x)-j\hbar\omega_\perp$ \cite{NarSta98}. Within this
model, the linear density is given by
\begin{equation}
n_{1}(x,\mu ,T)=n_{YY}(\mu (x),T)+\sum\nolimits_{j=1}^{\infty
}(j+1)n_{e}(\mu _{j}(x),T).  \nonumber  \label{eq:n1}
\end{equation}
For the radially excited states, we use the result of the LDA for
the 1D ideal gas, $n_e(\mu_j,T)=g_{1/2}(\exp(\mu_j/k_BT))/\Lambda_T$
where $g_{1/2}$ is a Bose function and $\Lambda_T=(2 \pi \hbar^{2}/m
k_B T)^{1/2}$ is the thermal de Broglie wavelength
\cite{PetGanShl04,BouKheShl07}. Note that as long as
$\mu<\hbar\omega_\perp$ we have $\mu_j<0$ which is necessary to
avoid divergence of $g_{1/2}$. In this model, the radially excited
states act as a bath for particle and energy exchange with the
radial ground state. The resulting fits are shown as solid lines in
Fig.~\ref {fig:SituFocus}(a)-(d) and describe our data very well.
The fitted values of $T$ and $\mu$ are displayed in
Fig.~\ref{fig:temperature}.

We now turn to the \emph{in focus} measurements which
give access to the axial momentum distribution of the gas.
The focusing pulse is created by ramping up the axial trapping
frequency from 8.5~Hz to 20~Hz in 0.8~ms, maintaining this for
3.8~ms, and ramping back to 8.5~Hz in 0.8~ms, followed by a sudden
switch-off of the magnetic trap. During the focusing pulse the cloud
length reduces by less than $20\%$. After switching off the magnetic
trap, the cloud expands in the radial direction on a timescale of
$1/\omega_\perp$, so that the interactions vanish rapidly compared
to the relevant axial timescale and the subsequent axial contraction
can be treated as free propagation. After 13~ms of free propagation
the cloud comes to a focus.

In Fig.~\ref{fig:SituFocus}(e)-(h) we show the axial density
distribution obtained in the focus, averaged over typically $10$
shots, for final RF frequencies similar to the \textit{in situ} data
in Fig.~\ref{fig:SituFocus} (a)-(d). Here, in contrast to the
\textit{in situ} results, one can clearly distinguish a narrow peak
from a broad pedestal for RF values below $1.56~$ MHz
[Figs.~\ref{fig:SituFocus}(g)-(h)]. The Yang-Yang solution does not
yield the momentum distribution and thus it can not be used to fit
to the \textit{in focus} data. Instead, to quantify the observation
of the bimodal structure we first fit a 2D Gaussian to the wings of
the atomic density distribution. In a second step we fit a narrow
Gaussian to the residual peak in the center. The fitted curves are
shown after integration in the $z$ direction in
Fig.~\ref{fig:SituFocus}(e)-(h), and describe the observed {\em in
focus} distributions well. Fig.~\ref{fig:temperature}(c) shows the
resulting atom numbers in the wide and narrow part of the momentum
distribution; we also plot the atom numbers from the Yang-Yang model
in the radial ground state, in the radially excited states, and
atoms in the radial ground state experiencing $\mu(x)>0$. Comparing
the {\em in situ} and the {\em in focus} data, we conclude that:
\emph{(i)} the momentum distribution becomes bimodal around the
point where the global chemical potential $\mu$ crosses zero and
becomes positive; and \emph{(ii)} the narrow part of the momentum
distribution is dominated by the atoms in the radial ground state
(described by $n_{YY}$), while the wide part is dominated by atoms
in the radially excited states. The shape of the narrow peak thus
constitutes a measurement of the momentum distribution of the
Yang-Yang gas, for which currently no theoretical comparison is
available.

A further comparison between the {\em in focus} and {\em in situ}
results can be made as follows. Estimates for the temperature can be
obtained from the Gaussian fit to the wide part of the {\em in focus}
data, by assuming that the tails (where degeneracy is negligible) are
well described by Boltzmann statistics. The resulting temperatures are
shown in Fig.~\ref{fig:temperature}(a). The agreement with the
temperature extracted from the {\em in situ} data is reasonable. We
attribute the remaining discrepancy to the approximations implicit in
the above interpretation of the Gaussian fit results, which neglects
the discrete radial level structure and the contribution of the radial
ground state to the wide part of the axial momentum distribution.

The failure of the ideal-gas and quasi-condensate descriptions is
illustrated in Fig.~\ref{fig:SituFocus}(c, d). The key point here
is the following. The Yang-Yang thermodynamic equations yield a smooth
equation of state $\nyy(\mu,T)$, including the region around
$\mu(x)=0$. This deviates dramatically from both the ideal-gas
description (diverging density as $\mu\rightarrow 0$ from below) and
the quasi-condensate description (vanishing density as $\mu\rightarrow 0$ from above). 
The region in $\mu(x)$ (and consequently in
$\nyy(x)$) where this discrepancy is significant \cite{BouKheShl07} is
particularly large for our parameters, and the Yang-Yang thermodynamic
solutions are essential for a proper description of the data.

In conclusion, we have found excellent agreement between \emph{in
situ} measurements of the spatial linear density of a nearly 1D
trapped Bose gas and a model based on the exact Yang-Yang solutions.
We have measured the corresponding momentum distribution, and expect
this to stimulate further theoretical work. More generally, our
results establish a new and strong link between experiments on
low-dimensional quantum gases and exact theory for interacting
quantum many-body systems at finite temperatures, both fields of
significant current interest. Additionally, our findings should be
relevant to related experimental systems, such as guided-wave
atom lasers \cite{GueRioGae06} and atom-chip based interferometers
\cite{HofLesFis07}.

We thank J.T.M. Walraven, G.V. Shlyapnikov and J.-S. Caux for fruitful
discussions. The atom chip was produced at the Amsterdam nanoCenter.
This work was funded by FOM, NWO and by
the EU (MRTN-CT-2003-505032). KK acknowledges support by the
Australian Research Council, the Queensland State Government, and
the Institut Henri Poincar\'{e} -- Centre Emile Borel.

\end{document}